\PassOptionsToPackage{dvipsnames}{xcolor}

\documentclass[manuscript]{aastex631}

\usepackage{amsmath}
\usepackage{graphicx}
\usepackage{CJK}
\usepackage{csvsimple}
\usepackage{comment}
\usepackage{soul}

\usepackage{placeins}

\usepackage{hyperref}
\usepackage{soul}
\usepackage{array}
\usepackage{multirow}
\usepackage{numprint}

\begin{document}

\title{Analyzing the Morphology of Late-phase Stellar Flares From G-, K-, and M-type Stars}

\author[0009-0008-0072-120X]{Denise G. Yudovich}
\affiliation{Department of Astronomy, University of Florida, 211 Bryant Space Science Center P.O. Box 112055, Gainesville, FL 32611-2055 USA}
\affiliation{Institute for Astronomy, University of Hawai`i at M\=anoa, 2680 Woodlawn Drive, Honolulu, HI 96822, USA}

\author[0000-0002-7663-7652]{Kai E. Yang}
\affiliation{Institute for Astronomy, University of Hawai`i at M\=anoa, 2680 Woodlawn Drive, Honolulu, HI 96822, USA}

\author[0000-0003-4043-616X]{Xudong Sun}
\affiliation{Institute for Astronomy, University of Hawai`i at M\=anoa, 2680 Woodlawn Drive, Honolulu, HI 96822, USA}

\correspondingauthor{Denise G. Yudovich}
\email{deniseyudo@gmail.com}


\begin{abstract}
Stellar flares occasionally present a \textit{peak-bump} light curve morphology, consisting of an initial impulsive phase followed by a gradual late phase. Analyzing this specific morphology can uncover the underlying physics of stellar flare dynamics, particularly the plasma heating-evaporation-condensation process. While previous studies have mainly examined peak-bump occurrences on M-dwarfs, this report extends the investigation to G-, K-, and M-type stars. We utilize the flare catalog published by \cite{Crowley_et_al._2022}, encompassing 12,597 flares, detected by using \textit{Transiting Exoplanet Survey Satellite} (TESS) observations. Our analysis identifies 10,142 flares with discernible classical and complex morphology, of which 197 ($\sim1.9\%$) exhibit the peak-bump feature. We delve into the statistical properties of these TESS late-phase flares, noting that both the amplitude and full-width-half-maximum (FWHM) duration of both the peaks and bumps show positive correlations across all source-star spectral types, following a power law with indices 0.69 $\pm$ 0.09 and 1.0 $\pm$ 0.15, respectively. Additionally, a negative correlation between flare amplitude and the effective temperature of their host stars is observed. Compared to the other flares in our sample, peak-bump flares tend to have larger and longer initial peak amplitudes and FWHM durations, and possess energies ranging from $10^{31}$ to $10^{36}$~erg.


\end{abstract}


\section{Introduction} \label{sec:intro}

Stellar superflares, flares being ``super" bolometric in their peak luminosities, present a unique and important avenue for exploring flare dynamics in the high-energy range, especially for understanding the stellar atmosphere's response to such a huge amount of energy \citep{Osten_et_al._2007}. Particularly, the study of superflares provides valuable insight into phenomena that greatly impact Earth's space weather if similar events were to occur on our Sun, i.e., the Carrington event that occurred in 1859 \citep{Hudson2021ARAA}. Moreover, these highly energetic superflares could significantly influence the habitability of exoplanets. Notably, they may \textit{initiate} vital chemical reactions crucial for the formation of life \citep{Airapetian_et_al._2016}, or accompanied by stellar coronal mass ejections (CMEs), energetic protons can lead to the depletion of ozone layers on magnetically unprotected Earth-like planets, though few observations have been reported for the possible existence of stellar CMEs \citep{Tilley_et_al._2019,Namekata2022NatAs...6..241N,Notsu2024ApJ...961..189N}. 

Stellar flares are widely believed to arise from magnetic field reconnection in their atmospheres \citep[][and references therein]{Usoskin2023SSRv..219...73U,kowalski_2024}. This reconnection process occurs when the magnetic field becomes anti-parallel or highly sheared. However, the exact conditions leading to the triggering of magnetic reconnection remain unclear \citep[][and references therein]{Pontin2022}. As a result of this reconnection, magnetic energy is released in various forms, including thermal energy, plasma bulk kinetic energy, high-energy particles, waves, and radiation. Flares are typically thought to occur in starspot-active regions where the magnetic field is particularly strong \citep[from 2,000-4,000 G at their central umbrae,][]{Murray2013,Oloketuyi_et_al._2023}. 
Typical solar flare energy ranges from $10^{28}$ to $10^{32}$ erg within a short time period, from several minutes to hours,
while stellar ``superflares" could release much larger energy \citep[$>10^{33}$ erg,][]{Schaefer_et_al._2000}. \cite{Maehara_et_al._2015} investigate 187 superflares on 23 G-type stars and show that their bolometric energies range from $10^{33}$--$10^{36}$ erg.

Since the launch of \textit{Kepler} \citep{Borucki2010} and \textit{Transiting Exoplanet Survey Satellite} (TESS, \citealt{Ricker_et_al._2014}), there has been a remarkable increase in the number of observed stellar flares \citep{Maehara2012,Balona2015,Davenport2016,VanDoorsselaere2017,Yang2019,Notsu2019,Gunther2020,Tu2020,Tu2021,Crowley_et_al._2024}. The majority of these stellar flares exhibit a single impulsive peak morphology in their light curves \citep{Davenport_et_al._2014}. Hereafter, we use the designation of ``classical flares" to refer to such flares. Stellar light curves can also show complex morphology that deviates significantly from this classical template, as reported by early high-cadence observations \citep{Moffett_et_al._1974, Bopp&Moffett}. More recent examples can be found in a statistical study conducted by \cite{Howard_&_MacGregor} which analyzed TESS flares following a complex morphology on M-dwarfs. \cite{Pietras2022ApJ} also found that a second gradual component, when present, dominates a flare's decay phase. Complex flares are typically composed of multiple peaks, which can be further categorized into three main types: peak-bump flares, flat-top flares, and Quasi-Periodic Pulsations (QPPs, \citealt{Zimovets2021,Ramsay2021SoPh}). Flat-top flares exhibit nearly constant emission at their peaks and are generally characterized as low-energy, low-impulse events. QPPs are characterized as those with small periodic bumps in the rise or decay phases. Lastly, peak-bump flares, the primary focus of this study, consist of a single impulsive peak followed by a secondary gradual bump. Oftentimes, the initial peak displays a higher amplitude while the subsequent bump has a smaller amplitude. The bump, however, can dominate at late times during the decay phase of the original peak \citep{Pietras2022ApJ}. \cite{Howard_&_MacGregor} suggested that peak-bump flares constitute approximately $7\%$ of all 440 flares in their sample of M-dwarfs. We note that peak-bump flares were readily reported by high-resolution observations as ``secondary flares" in works prior to TESS. In particular, \cite{kowalski_et_al._2016} analyzed a large sample of M Dwarf flares from ULTRACAM at 1.5s cadence. Their analysis of a flare event reveals a complex secondary phase in the light curve, (marked as F2 in their samples). In addition, peak-bump flares have been reported in high-resolution data from ground-based observatories and XMM-Newton in different wavelengths \citep{Fuhrmeister_et_al._2011, Garcia_et_al._2002, hawley&pettersen1991, kowalski2013}. 

The cause of flare complexity in the optical wavelength remains generally uncertain, however, there are two main hypotheses: the flare-cascade scenario and sympathetic flaring \citep{Davenport_et_al._2014,Hawley_et_al._2014}. The flare-cascade scenario proposes that subsequent flares from the same active region decrease in amplitude over time, leading to complex multi-peak morphology. Sympathetic flaring, in contrast, implies distinct but interconnected source regions. An initial eruption can trigger a second eruption in a nearby magnetic structure through magnetic interactions \citep{Lynch_et_al._2016}.
 While sympathetic flaring has been observed on the Sun \citep{Moon2002ApJ,Schrijver_&_Higgins_2015}, further research is needed to understand the underlying mechanism behind this phenomenon. \cite{Howard_&_MacGregor} speculate that the peak-bump structure in TESS observations from M-dwarf flares can arise from an initial magnetic field reconnection, which generates a single impulsive peak, followed by a more gradual release of energy manifesting as a bump, a process that is currently not well understood.


Another possible source for the peak-bump flare emission is the dense, off-limb coronal loops. In solar observations, the peak-bump morphology is commonly observed in extreme ultraviolet (EUV) and soft X-ray (SXR) wavebands \citep{Woods2011, Qiu2012}. This phenomenon is referred to as the ``late-phase flare," and is typically attributed to multiple flare loops of varying lengths or multiple heating processes. Due to their similar light curve morphology, we will use the terms ``late-phase flare" and ``peak-bump flare" interchangeably hereafter. We note that the mechanism of the white-light flare emission is quite different from that of the EUV and SXR emissions. Most optical flare emission shows a single peak in time and mostly comes from the lower solar atmosphere instead of coronal loops, with rare exceptions \citep{Matthews2003,Kerr2014ApJ,Hao2017}. Nevertheless, recent observations from the Helioseismic and Magnetic Imager \citep[HMI,][]{Schou2012SoPh..275..229S,Scherrer2012SoPh..275..207S} onboard the Solar Dynamics Observatory \citep[SDO,][]{Pesnell2012SoPh..275....3P} have reported that off-limb post-flare loops are observed during the gradual phase of certain solar flares, which showed significant density enhancement \citep{Saint-Hilaire2014ApJ...786L..19S,MartinezOliveros2014ApJ...780L..28M,MartinezOliveros2022ApJ...936...56M,Jejcic2018,Zhao2021}. The most definitive case comes from the large solar flare \texttt{SOL2017-09-10T15:35}, which shows clear coronal emissions in the optical waveband. The flux from off-limb post-flare loops persists for nearly an hour, exhibiting a much more gradual morphology compared to that seen in EUV and SXR (see Figure 4 in \citealt{Zhao2021}). These solar events, as expected, are orders of magnitude less energetic than those observed in the optical band by TESS. 

These solar observations support the theory of \cite{Heinzel2017} and \cite{Heinzel2018} on the role of optically thin coronal emissions in enhancing white-light intensity during large flares, especially when plasma density ($n_e$) is $>10^{12}$--$10^{13}$ cm$^{-3}$. While the integrated signal of the Sun is small, it is spatially resolved. Hydrodynamic (HD) flare simulations by \cite{Yang2023} integrate these observations and theories, providing a comprehensive understanding of the dynamic process in the corona during the flare's bump phase. During the evolution of the flare loop, the chromospheric plasma heats up extensively, evaporates into the corona, and loses energy through optically thin emission. Once the plasma density becomes high enough in the corona, thermal instability is triggered \citep{Parker1953,Field1965,Claes2020}, causing the coronal plasma to compress and cool significantly before falling back to the solar surface, which is known as \textit{coronal rain} \citep{Scullion2016,Antolin2022,Mason2022}. 
These loops' optical emissions mainly come from the flare-induced coronal rain, which is composed of highly condensed cool plasma, and typically form with a time delay following the initial heating. Cooling times observed in EUV and SXR flare loops vary significantly, ranging from a few to tens of minutes \citep{ryan_2016}, and can differ between different loops due to variations in loop length and thermal dynamics. Such delayed emission from the coronal rain can potentially explain the distinct secondary bump in TESS peak-bump flares. Indeed, the observed temporal offsets from the impulsive peak match the cooling time scales from the HD simulations \citep{Yang2023}. 

We note that \cite{Yang2023}'s hypothesis, i.e., coronal rain causing the white-light late phase, is just one of many. Another possible mechanism rests on the idea that complex magnetic configurations can lead to a multitude of reconnection events in the same active region. For example, \cite{Kosovichev&Zharkova}, analyzed the ``Bastille Day Flare" event and found that the flare ribbons evolved into different spatial regions throughout the active region. Different ribbons ignited at different times with offsets of about 10 minutes; the integrated light curve may show a delayed emissions peak as observed in TESS late-phase flares. \cite{Kowalski2017ApJ...837..125K} demonstrated that the secondary peak can be created by newly heated flare loops, instead of the cooling of loops heated during the impulsive phase.


Our research focuses on expanding the understanding of peak-bump stellar flares on cool stars, to improve constraints on high-energy flare theories. While the peak-bump structure has been thoroughly studied for M-dwarfs \citep{Howard_&_MacGregor}, its properties remain largely unexplored for other cool stars. In this paper, we aim to identify the presence of the peak-bump morphology on G-, K-, and M-dwarfs and investigate the statistical relationship between its two components: the peak and the bump. The dynamics of stellar flares are intricately linked to the host star's magnetic field, which is influenced by rotation rate \citep{goodarzi_2019}. Stars of different effective temperatures exhibit variations in rotation rates \citep{McQuillan2014}, surface gravity, and atmospheric composition, all of which shape the flare behavior. The statistical properties of peak-bump flares can shed light on the relationship between stellar properties and flare behavior across diverse stellar environments. Although we briefly discuss the presence of late-phase flares in a small sample of F-type stars to demonstrate that such complex flares can occur across a broad spectral range, our primary focus lies in investigating these flares on G-, K-, and M-dwarfs.

This paper is arranged as follows. Section \ref{s.method2} outlines the data obtained from \textit{TESS} and our approach to generating light curves. In Section \ref{s.method3}, we explain the method used to identify peak-bumps and examine their morphology. Section \ref{s.method4} presents a separate comparison of the peak and bump features. Finally, in Section \ref{s.method5}, we provide a comprehensive summary of our findings along with a discussion of the proposed physics behind such kinds of stellar flares.                                             


\vspace*{-0.7cm}

\section{TESS Data and Flare Sample} \label{s.method2}
The TESS mission primarily aims to discover transiting exoplanets by using four 10.5 cm telescopes to gather data on photometric observations in the optical bandpass (600--1000 nm). Many stellar flares have been reported by TESS's high-quality observations. Analysis of M-dwarf flares with 20-second versus 2-minute cadence data shows significant structural differences and reveals complex substructures in nearly half of the observed large flares \citep{Howard_&_MacGregor}. 
High cadence data have been shown to reduce degeneracy in interpretation and offer more precise information on stellar flare morphology, both of which are crucial for understanding flare emissions on short time scales (from seconds to tens of seconds). This research leverages TESS's 20-second cadence data for an in-depth analysis of flares from cool stars.

\begin{figure}
    \centering
    \includegraphics[width=0.75\linewidth]{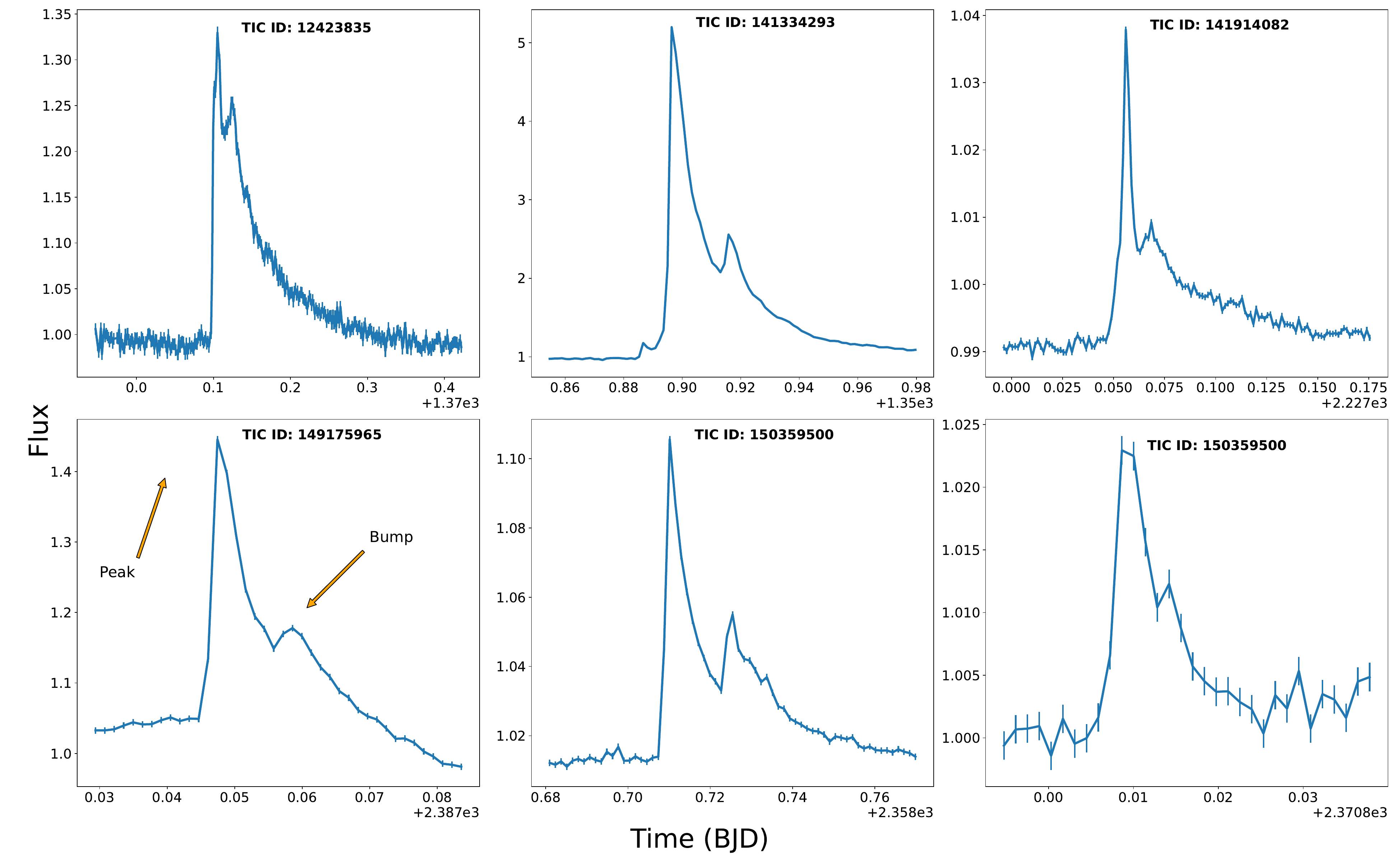}
    \caption{Examples of some visually-identified peak-bump flares in our sample. These late-phase flares are characterized by an initial impulsive peak followed by a secondary gradual bump. }
    \label{Figure 1.}
\end{figure}

Our flare sample was extracted from the flare catalog presented in \cite{Crowley_et_al._2022}, consisting of 12,597 flares from 234 active stars. This dataset incorporates their published catalog of G-type star flares and additional unpublished data on flares from F-, K-, and M-type stars, which we obtained through personal communication with the authors. Here we use the following temperature definitions: 6000-7500 K for F-type stars, 5000-6000 K for G-type stars, 3500-5000 K for K-type stars, and 2800-3500 K for M-dwarfs. Because our sample of flares from F-type stars is small, we will focus our analysis on G-, K-, and M-dwarfs. The light curves were carefully detrended for astrophysical modulation, using the algorithm from \cite{Chang2015ApJ}. We focused on flares identified with over 75\% confidence based on their amplitude-FWHM-recovery test, and further reduced this sample to 10,142 flares using the criteria described in Section \ref{s.method3}. Then, we visually identified a subsample of peak-bump flares, thus showing their occurrence on different types of stars (late-phase flares were already shown to exist on M-dwarfs; see \citealt{Howard_&_MacGregor}). In Figure \ref{Figure 1.}, we show examples of some light curves of identified peak-bump flares. 



\section{Flare Characterization} 
\label{s.method3}

Stellar flares exhibit a wide range of morphologies, as shown in previous observations \citep{hawley&pettersen1991, Garcia_et_al._2002, kowalski2013, kowalski_et_al._2016}. In this study, we primarily adopt the morphology classification outlined by \cite{Howard_&_MacGregor}. Accordingly, we adopt a methodology that involves fitting the light curves with distinct flare templates and assessing the quality of these fits to determine the type of flares in our sample.
We apply seven flare templates across our sample of 12,597 flares. Specifically, three of these templates are designed for modeling peak-bump flares, two are used for the classical flare type morphologies, and the remaining two, i.e.~QPP and Flat-top, are each represented by a single template. 

Below, we outline the templates for classical and peak-bump flare morphologies used for classifying flares, with the primary intention of isolating peak-bumps. Templates for other flare morphologies are described in Appendix \ref{app:1}, i.e.~the ``QPP Model" and ``Flat-top Model."
For classical flares, characterized by an impulsive increase followed by a slow decay, we use the template from \cite{Davenport_et_al._2014}, hereafter, this template will be referred to as the ``Classical Model." We also implement a ``Duo-Classical'' Model, defined as the overlay of two classical flare models, to differentiate between single flare events and events with multiple flares close in proximity. The flares best fit by this model are discarded in our subsequent analysis. 




Since there exists no available physics-based template for the bump phase of ``peak-bump" flares, we consider three empirical functions, the Chi-squared distribution ($\mathrm{F_{\chi^2}}$), Log-normal distribution ($\mathrm{F_{Lognorm}}$), and Gaussian/Norm distribution ($\mathrm{F_{Gaussian}}$), to represent the bump component. 
They are formulated as follows: 
\begin{equation}
\begin{split}
\mathrm{F_{\chi^2}} &= \dfrac{1}{2^{k/2}\Gamma(k/2)}\left(\frac{t-a}{w}\right)^{k/2-1}\exp \left(\dfrac{a - t}{2w}\right),  \\
\end{split}
\end{equation}
\begin{equation}
\begin{split}
\mathrm{F_{Lognorm}} &= \frac{w}{k(t-a)\sqrt{2\pi}}\exp \left(-\frac{\log^2{((t-a)/w)}}{2k^2}\right),\\
\end{split}
\end{equation}
\begin{equation}
\begin{split}
\mathrm{F_{Gaussian}} &= \exp \left(\frac{-((t-a)/w)^2}{2}\right) \frac{1}{\sqrt{2\pi}},
\end{split}
\end{equation}
where $k$ represents degrees of freedom, $a$ is the flare start time, $w$ is the flare width, and $t$ is the time.
These three templates are linearly superposed with a classical flare template to form three distinct peak-bump flare models, effectively capturing both the impulsive and bump phases of the peak-bump flare light curve. The use of multiple peak-bump flare templates allows for flexibility in the case that a particular model fails to adequately fit the data.

We use the Levenberg-Marquardt algorithm in the fit function \texttt{scipy}.\texttt{optimize}.\texttt{curve\_fit} to model each observed flare light curve with the aforementioned flare templates. Then, the qualities of the fits for all templates are measured by using the Akaike Information Criterion (AIC, \citealt{Wei2001}) and Bayesian Information Criterion (BIC, \citealt{kass1995reference}), which are defined as follows:
\begin{equation}
\mathrm{AIC} = N \, \log \frac{SS_{e}}{N} + 2k,
\end{equation}
\begin{equation}
\mathrm{BIC} = N \, \log\frac{SS_{e}}{N} + k \log N
\end{equation}
where $k$ is the number of parameters in each model, $SS_{e}$ is the sum of squared residuals between the model predictions and the observed data, and $N$ is the data sample size \citep{Akaike1974,Wit_et_al._2012}. These two statistical criteria evaluate how well the models match the light curves, considering the degrees of freedom for each model, and providing the means for model selection. 
We utilize both criteria to aggressively tackle the flare classification procedure, acknowledging that BIC fails to provide as good a fit as AIC for more complex data, and that AIC is overcome by BIC in situations with relatively simple data \citep{Aho_et_al._2014}. This procedure ensures a robust and comprehensive automatic classification of flares in our catalog. Details of the criteria for the morphology classification process are described in Appendix \ref{app:1} and \ref{app:2}. 
It is important to note that we discarded 773 flares from our initial sample of 12,597 due to insufficient light curve data and unsuccessful fit.
Considering the reduction from the ``Duo-Classical" Model, the final sample contains 10,142 flares. 
\begin{deluxetable*}{ccccccc}[t]
\centering
\caption{\label{Table 1.} Flare Classification Results}

\tablehead{
\colhead{ } & \colhead{Classical} & \colhead{Flat-top} & \colhead{QPP} & \colhead{Peak-bump} & \colhead{Complex}
}
\startdata
Flare Quantity & 3985 & 1277 & 95 & 197 & 6157 \\
Flare Percentage & 39\% & 12.6\% & 0.9\% & 1.9\% & 61\% 
\enddata
\bigskip
\textbf{Notes.} Shown are the results of our flare classification procedure. Note that the ``Complex" category includes  flat-top, QPP, and peak-bump flares, as well as the flares without any rigid characterization. 
\end{deluxetable*}

\begin{figure}
    \centering
    \includegraphics[width=0.75\linewidth]{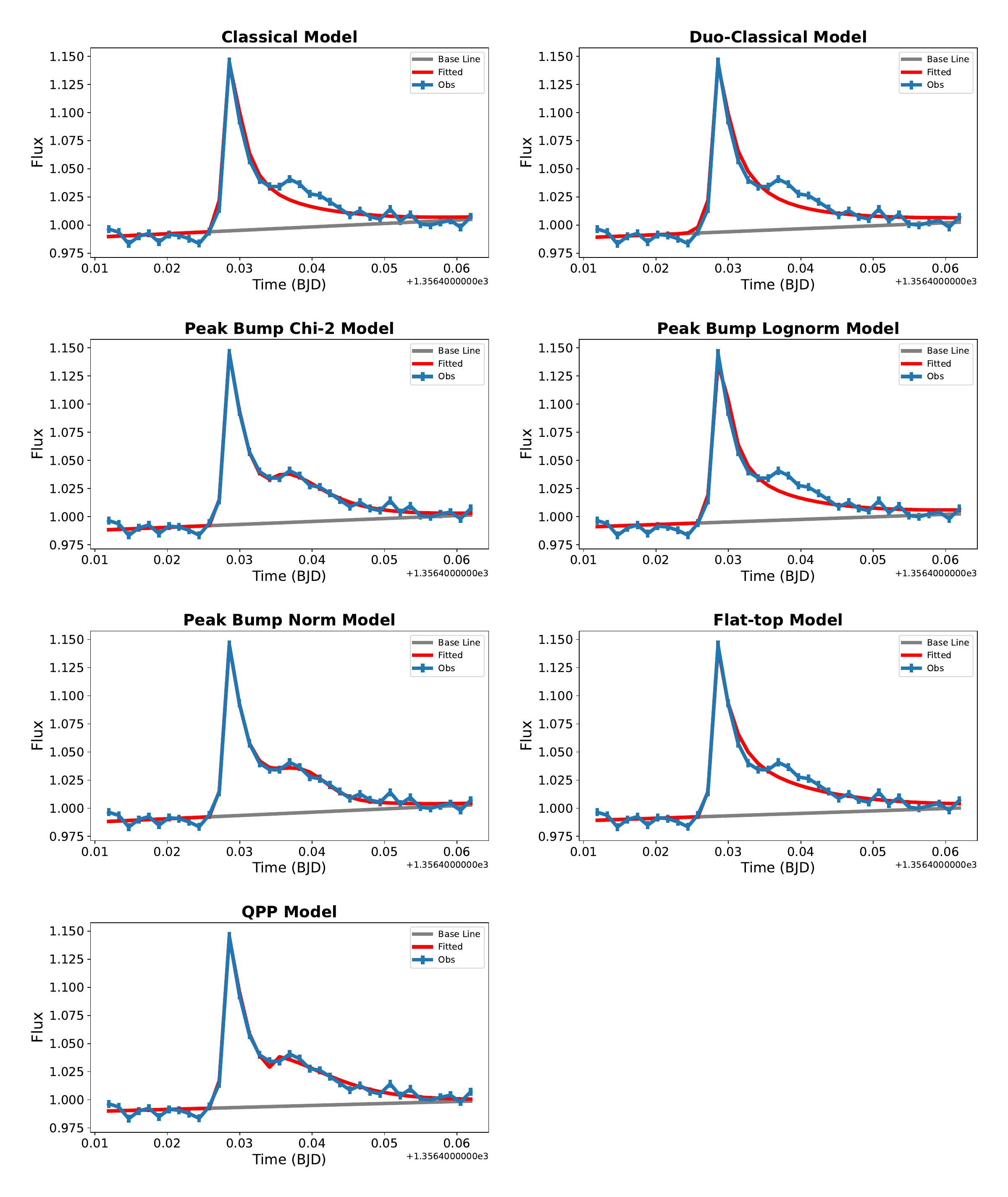}
    \caption{Example of a complex flare, specifically peak-bump, identified via our algorithm. Here seven fitting models are used to fit the flare. Each plot displays the flare's light curve (with the observed data in blue) and each model's resultant fit, marked with a red curve. Due to the secondary bump in the light curve, both the AIC and BIC criteria agree that this flare is best fit by the Peak Bump Chi-squared Model; this flare is thus complex late-phase. }
    \label{Figure 2.}
\end{figure}

In Table \ref{Table 1.}, we outline the results of the fitting and model comparison processes for the flares in our sample of 10,142 flares. The fitted features of some flares in this sample are listed in Table \ref{Table 2.}.
Figure \ref{Figure 11.} in Appendix \ref{app:2} shows an example of a classical flare fit by all seven models, emphasizing that the best fit is provided by the Classical Model, as agreed upon by both the AIC and BIC criteria. There are 3,985 classical flares in our sample, $39\%$ of the 10,142 flares in our renewed sample. The rest are 6,157 complex flares, $61\%$ of the sample. These include the flat-tops, peak-bumps, and QPPs, as well as cases that are not well fit by any models. Our focus is on the 1,132 peak-bump flares detected by this automatic procedure, which account for around $11\%$ of the flares in our reduced sample. Figure \ref{Figure 2.} shows an example of a peak-bump flare fit by all seven models, emphasizing that the best fit is provided by the Peak Bump Model. Detailed analyses of the other types will be deferred to future work.

\begin{figure}
    \centering
    \includegraphics[width=0.75\linewidth]{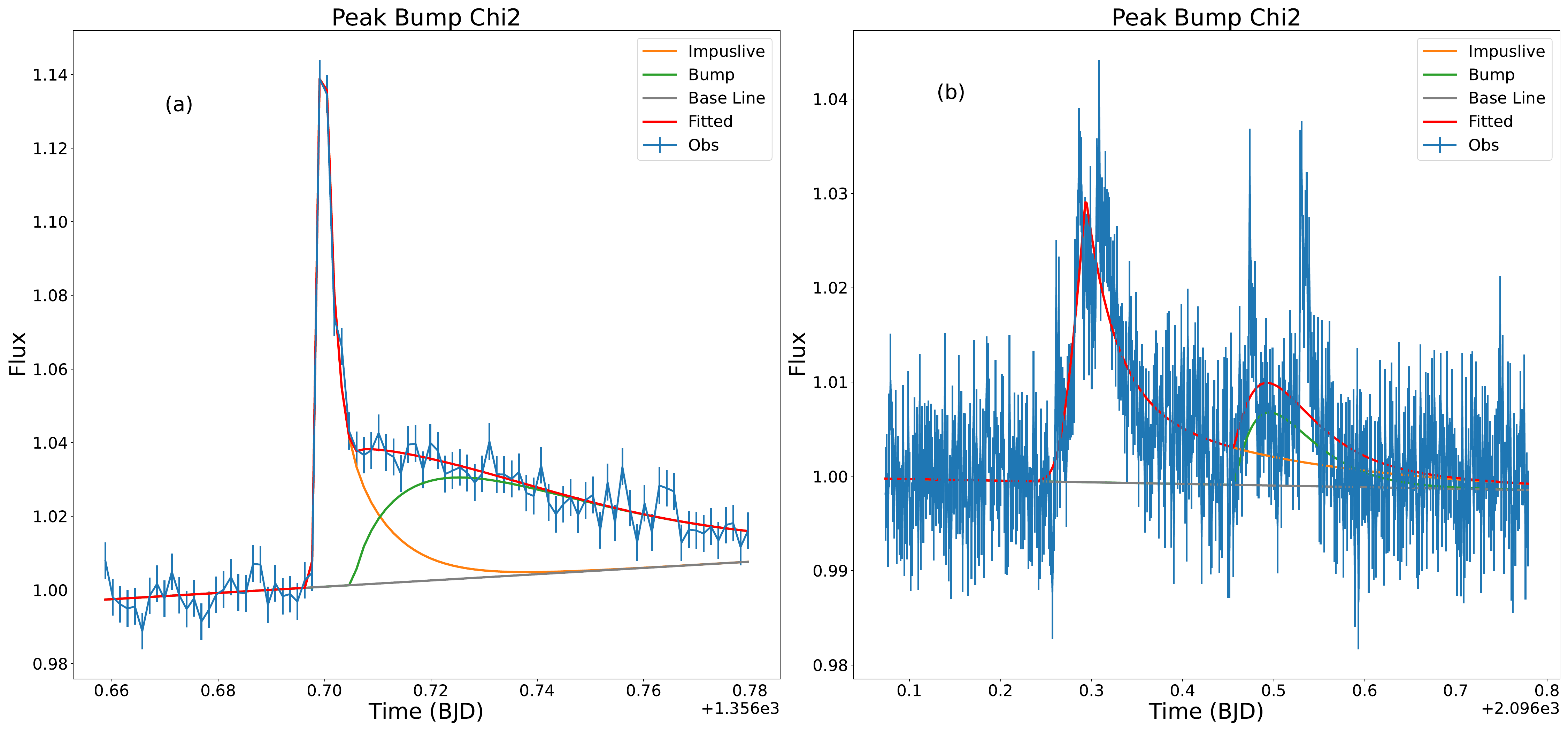}
    \caption{Examples of some flares that pass the statistical tests outlined in Section \ref{s.method3} but fail visual inspection. Thus, these flares are NOT included in our final ``true" peak-bump sample. Both flares are best fit by the Peak-bump Chi-squared model as determined by both the AIC and BIC criteria. However, the flare in panel (a) has an extended decay phase, rather than a singular post-impulse bump, and the flare in panel (b) has multiple ``secondary flares" following the impulsive phase, rather than a single smooth bump.}
    \label{Figure 3.}
\end{figure}

We have carried out additional tests to create a cleaner sample of peak-bump flares with minimal false positives. For the peak-bump sample selected from the previous steps, we first calculate the BIC difference between the best and second-best fitted models. A difference $\Delta \text{BIC} < 8$ suggests that there is no strong evidence to prefer one model over the other \citep{kass1995reference}, and the flare will be excluded from the sample. We then assess the goodness of fit by computing the $p$-value for the Anderson-Darling one-sample test \citep{Anderson10.1214aoms1177729437}. If the $p$-value does not exceed 0.01 (i.e. $p$ $<$ 0.01), we conclude that our flare templates do not fit the data well, and the flare will be excluded from the sample. These two tests reduce the late-phase flare sample to 652 flares. Finally, we visually inspect these light curves and exclude those with a morphology that deviates from the definition described in the Introduction. Examples of flares that pass the statistical tests but are finally excluded upon visual inspection are shown in Figure \ref{Figure 3.}. The fitting results of both these flares reveal a ``bump" phase that is actually an extended decay phase (panel (a)) or that combines two potential ``bumps" into one (panel (b)). Discarded events are designated as ``Complex" instead. Within the original peak-bump sample, we find that 197 of these flares are ``true" peak-bump flares that pass these statistical and visual tests. This approach results in a conservative estimate of the peak-bump flare occurence rate over our total flare sample.

\begin{figure}[t]
    \centering
    \includegraphics[width=1\linewidth]{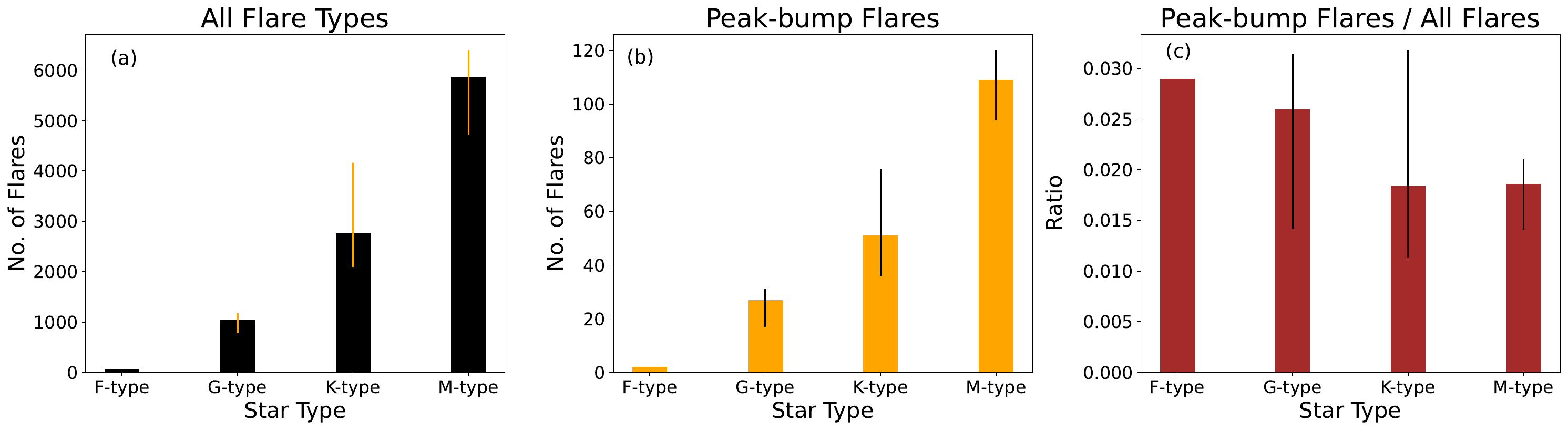}
    \caption{Bar plots indicating the number of flares from F-, G-, K-, and M-type stars in our sample. Panel (a) displays the total number of flares from the corresponding spectral type, while the orange bars in panel (b) represent
 the total number of detected peak-bump flares from each star type. Panel (c) displays the ratios of the detected peak-bump flares to total flares for each spectral type. The vertical lines represent the corresponding uncertainties in the number of flares}
    
    \label{Figure 4.}
\end{figure}

For this clean sample of late-phase flares, we find that M-dwarfs contribute the highest number of peak-bumps, $109\substack{+11 \\ -15}$ flares (around 55\% of the sample), nearly double the amount produced from K-type stars and over four times that produced from G-type stars, $51\substack{+25 \\ -15}$ flares and $27\substack{+4 \\ -10}$ flares (26\% and 14\%), respectively. The uncertainty values of the flare number were calculated given the temperature uncertainty of the source star. The different flare numbers may be attributed to the flare sample size for each spectral type, as the M-dwarf category has the largest flare sample, and thus the probability of finding a peak-bump flare is greater. We note that this result is consistent with \cite{Candelaresi_et_al_2014}'s conclusion that stars with lower effective temperatures produce more frequent superflares. Additionally, Figure \ref{Figure 4.} illustrates this data (in panel (b)) compared to the total number of flares produced from each stellar type (panel (a)). With respect to sample size, G-type stars seem to produce a higher percentage of peak-bump flares ($2.60\%\substack{+0.54\% \\ -1.18\%}$) than K and M-type stars ($1.85\%\substack{+1.33\% \\ -0.72\%}$ and $1.86\%\substack{+0.25\% \\ -0.45\%}$, respectively) as displayed in Figure \ref{Figure 4.} panel (c). To note, our sample has three F-type stars contributing two peak-bump flares (1\%) to the total late-phase flare sample. These peak-bump flares make up $2.9\%$ of the total flares produced from F-type stars in our sample. It is worth noting that the observed decrease in occurrence rates across different stellar types remains within the uncertainties. Given the limited flare sample size, this trend is not robust enough to draw definitive conclusions, highlighting the need for further statistical analyses. Lastly, we don't have temperature information on the stars hosting the remaining flares in our late-phase flare sample and they are thus not represented in Figure \ref{Figure 4.}.



We also performed a separate analysis of the total 1,132 peak-bump flare sample using only visual inspection without the statistical tests described above. This yields a larger sample of 426 flares. The results are shown in Appendix \ref{app:3} and they support the main results detailed in Section \ref{s.method4}, which were produced from the 197 peak-bump flare sample. 

\vspace*{-0.2cm}

\begin{table}

\caption{\label{Table 2.} Catalog of the 10,142 Flares Observed in Our Sample}

\begin{tabular}{|l |l |l |l| l |} 
 \hline
 TIC-ID & Peak (BJD) & FWHM (Day) & Amplitude (\(\Delta{F}/F\)) & Classification \\ [0.5ex] 
 \hline
 101963752 & 2111.0979 & 0.0076 & 0.0417 & Other (Complex) \\ 
 102032397 & 1370.8371 & 0.0051 & 0.0349 & Classical \\
 102071750 & 1378.1649 & 0.0028 & 0.0921 & Flat-top \\
 141334293 & 2231.1958 & 0.0047 & 0.0277 & Classical \\
 114794572 & 1378.5262 & 0.0610 & 0.0682 & Peak-bump \\
 114851071 & 1380.2970 & 0.0076 & 0.0769 & Other (Complex) \\
 102032397 & 1368.9968 & 0.0051 & 0.0812 & QPP \\[1ex]
  \hline
\end{tabular}

\bigskip
\textbf{Notes.} Examples of some flare data from the 10,142 flares in our sample, extracted from \cite{Crowley_et_al._2022}. Listed, are the flares' peak times in Barycentric Julian Date (BJD), full width at half maximum (FWHM) durations in days, and amplitudes in fractional flux. The TESS ID of each flare's star is also listed, along with each flare's classification determined by our algorithm. The entirety of Table \ref{Table 2.} is published in the machine-readable format. 





\nolinenumbers
    \begin{center}
    \caption{\label{Table 3.} Catalog of 197 Detected Late-phase Flares}
    \vspace*{-5mm}
    \begin{tabular}{|p{2cm}|p{2cm}|p{2cm}|p{1.5cm}| p{2cm}|p{2cm}|p{1.8cm}|}  
    \hline 
         { TIC-ID}  &  { Peak (BJD)} &  { Amplitude ($\Delta F/F$)} &  { FWHM (Day)} &   {Bump Peak (BJD)}  &  { Bump FWHM (Day)}  & {Bump Amplitude  ($\Delta F/F$)} \\ \hline  
         12423835&  1370.1051 &  0.3252&  0.0318&  1370.1334&  4.08e-3& 0.2484\\   
         141334293&  1350.8964&  3.8242&  0.0085&  1350.9060&  9.530e-3 & 0.7956\\ 
         141807839&  2287.0773&  0.0916 &  0.0035 &  2287.0848&  2.404e-03& 0.0166\\   
         141807839&  1456.7146&  0.4345&  0.0093&  1456.7216&  3.502e-03& 0.1603\\ 
         141334293&  1356.6991&  0.1258&  0.0081&  1356.7238 &  1.264e-02& 0.0276\\  
         271900514&  2377.9783&  0.0667&  0.0212&  2378.0518&  3.063e-02& 0.0155\\ 
         278634010&  2082.7297 &  0.0067 &  0.0071&  2082.7491&  4.877e-04& 0.0024\\
         \hline  
    \end{tabular}
    \end{center}
    \textbf{Notes.} Examples of some peak-bump flare features (extracted from \cite{Crowley_et_al._2022}) within our narrowed-down sample of 197 flares. Listed are the flares' peak times in BJD, peak amplitudes in fractional flux (Amplitude $\Delta F/F$), peak FWHM durations in days, bump peak times in BJD, bump FWHM durations in days, bump amplitudes in fractional flux (Bump Amplitude $\Delta F/F$), and the source stars' TESS IDs. The entirety of Table \ref{Table 3.} is published in the machine-readable format. 
    
\end{table}

\clearpage

\vspace*{-1mm}
\section{Comparisons of Peak and Bump Features}\label{s.method4}

We list the fitted properties of peak-bump flares in Table \ref{Table 3.}. These include the flares' peak amplitudes (the values of maximum flux), FWHM durations (the width between two points at half-maximum amplitudes), and peak times (values at points of maximum flux) of the peak and bump components, respectively. 

\begin{figure}[ht]
    \centering
    \includegraphics[width=1\linewidth]{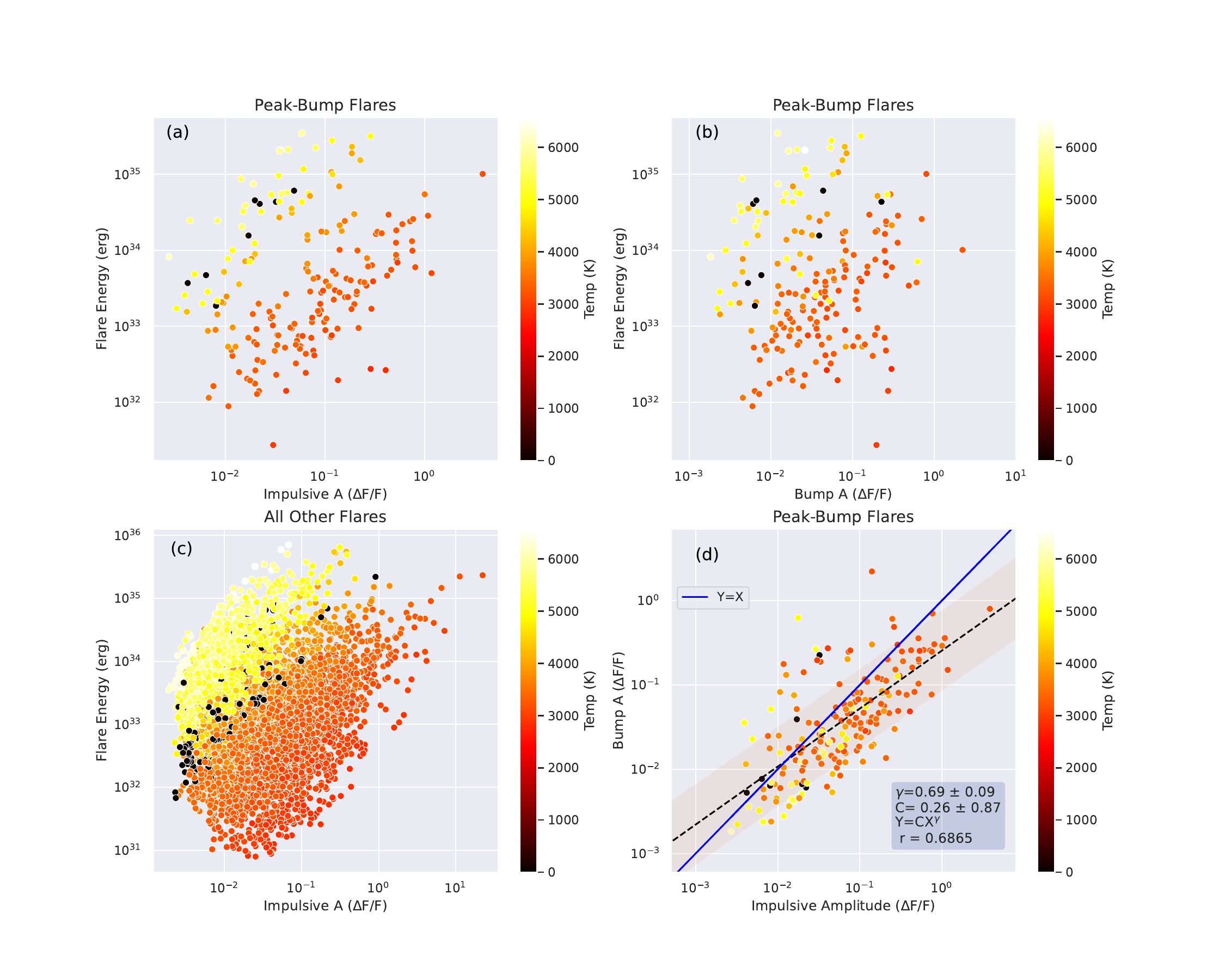}
    \caption{A comparison between the amplitudes of each fitted flare component for 197 peak-bump flares. Plot (a) shows a positive  relation between flare energy and the amplitude of the impulsive component. Plot (b) shows a positive  correlation between flare energy and the bump amplitude. Panel (c) shows a positive  relation between flare energy and the impulsive amplitude for all other flares in our sample. Lastly, panel (d) shows a positive  relation between the amplitudes of the impulsive and bump components. Overall, the color of each point indicates the effective temperature of the flare's source star from the TIC v8 catalog, while the black dots represent the stars with undetermined temperatures. The Spearman correlation coefficient of plot (d) is 0.69. The fitted power-law index of the impulsive-bump amplitude relation in plot (d) is 0.69 $\pm$ 0.09.}
    
    \label{Figure 5.}
\end{figure}

\begin{figure}[ht]
    \centering
    \includegraphics[width=1\linewidth]{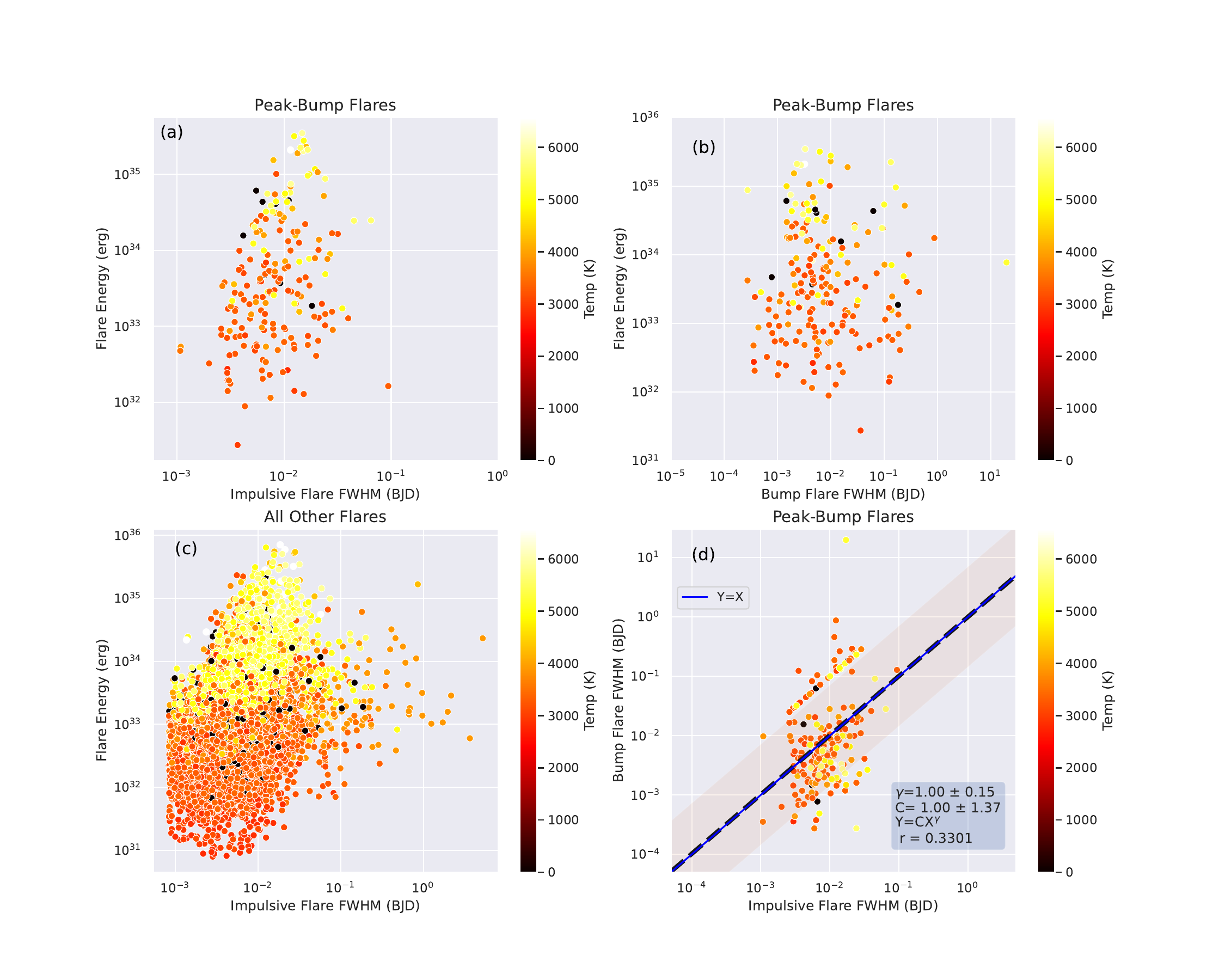}
    \caption{Same as Figure \ref{Figure 5.} but for the FWHM duration of each flare component for peak-bump flares. The Spearman correlation coefficient of panel (d) is 0.33, and the fitted power-law index of the relation in (d) is 1.00 $\pm$ 0.15. We note that the grouping of points above the fitted trendline represents the data of the peak-bump flares with longer FWHM durations for their bump components than those for the flares outside the grouping. }
     \label{Figure 6.}
\end{figure}

Figures \ref{Figure 5.} and \ref{Figure 6.} present scatter plots of the peak-bump features outlined in Table \ref{Table 3.}, and compare these features to those for all other flare types in our refined flare catalog. 
Specifically, Figure \ref{Figure 5.} shows positive  correlations between flare energy and impulsive and bump amplitudes, while Figure \ref{Figure 6.} displays those between flare energy, and impulsive and bump FWHM durations.
The flare energy values displayed in these plots are calculated using the relation presented in \cite{Crowley_et_al._2022}.

Figure \ref{Figure 5.}(a) and (b) illustrate the relationship between the energy of flares, 
and the amplitudes of the impulsive and bump components in peak-bump flares, respectively. 
These plots show roughly positive correlations, which indicate that with greater flare energy, both the impulsive and bump amplitudes of peak-bump flares increase. This amplitude relation agrees with \cite{Hawley_et_al._2014}'s work on multiple peak complex flares from the star GJ 1243. We note that in panel (a) there appears to be a gap for K-type stars. Although we are not certain what might be causing the gap, a reason might be the possible existence of a variable flare energy maximum limit for cooler K-stars and a variable flare energy minimum limit for hotter K-stars, within the impulsive amplitude range of about $0.01-0.1$ [$\Delta F/F$] . 
In Figure \ref{Figure 5.}(c), we show the  relation between flare energy and impulsive amplitude for all other flares in our sample; a similar positive  correlation is shown, implying that the peak-bump's amplitude-energy correlation is not unique to this flare type.
Surprisingly, Figure \ref{Figure 5.}(d) shows a relatively tight, positive  relation between the amplitudes of the impulsive and bump components. The Spearman correlation coefficient for this trend is 0.69 (calculated using the natural values, not the log values), and the fitted power-law index is 0.69 $\pm$ 0.09. This strong correlation supports the idea of the interconnectedness of the peak and bump profiles for such flares. It is important to note that this connection remains true for F-, G-, K-, and M-dwarfs. Additionally, the color of each point in all plots indicates the effective temperatures, obtained from the Tess Input Catalog \cite[TIC, v8,][]{Stassun2019AJ}, of the stars from which the flares are produced. In comparing effective temperatures and energies, Figure \ref{Figure 5.}, panels (a), (b) and (c), show a positive correlation between the two. This may be attributed to the simplified flare energy calculation method \citep[see Equations (7)--(9) in][]{Gunther2020, Hawley&Fisher_1992}, which relies on the idea that the flare energy is proportional to the stellar luminosity of the source star and that the flare temperature is a fixed constant, $9000$ K, differing from multi-wavelength observations \citep{Berger2024}. We note that \cite{YangZ._et_al._2023}'s work disagrees with the aforementioned correlation, however, this may be due to the differences in our methods of calculating flare energies. Figure \ref{Figure 5.} panels (a), (b), and (c), also highlight the discrepancy between flares from different stellar spectral types, particularly emphasizing that the higher temperature stars (F- and G-dwarfs) tend to have flares (of any type) with lower impulsive amplitudes (and lower bump amplitudes for peak-bump flares).

Figure \ref{Figure 6.} panels (a) and (b) also show a positive trend between flare energy and FWHM duration for both the impulsive and bump components of peak-bump flares. This result agrees with the energy-duration relation of multiple peak complex flares from the star GJ 1243 found in \cite{Hawley_et_al._2014}. A positive correlation is also seen in panel (c) for all other flare types in our sample. When comparing the peak-bump flares' amplitudes and FWHMs with their stars' effective temperatures, the amplitudes appear to be inversely proportional to the effective temperatures of the source stars, while the FWHMs lack a comparable relationship, as seen in Figures \ref{Figure 5.} and \ref{Figure 6.}. This amplitude-temperature correlation may stem from an observational bias in which flares are not only more readily detected on less luminous stars, but their amplitude ranges are also more pronounced. Figure \ref{Figure 6.} panel (d) also displays a positive relation between impulsive and bump FWHM durations, furthering the idea that the two profiles are connected. The Spearman correlation coefficient for this trend is 0.33 (calculated using the natural values, not the log values), and the fitted power-law index is 1.0 $\pm$ 0.15 (likely dominated by the narrow grouping of points above the identity line). Additionally, above the trend line in panel (d), there exists a grouping of data points. This grouping marks the peak-bump flares with similar FWHM durations for their impulsive components, albeit much longer FWHM durations (by about an order of magnitude) for their bump components when compared to those outside the grouping. These flares still follow the same trend as the flares outside the grouping, and are in fact more in tune with it. 

\begin{figure}
    \centering
    \includegraphics[width=0.75\linewidth]{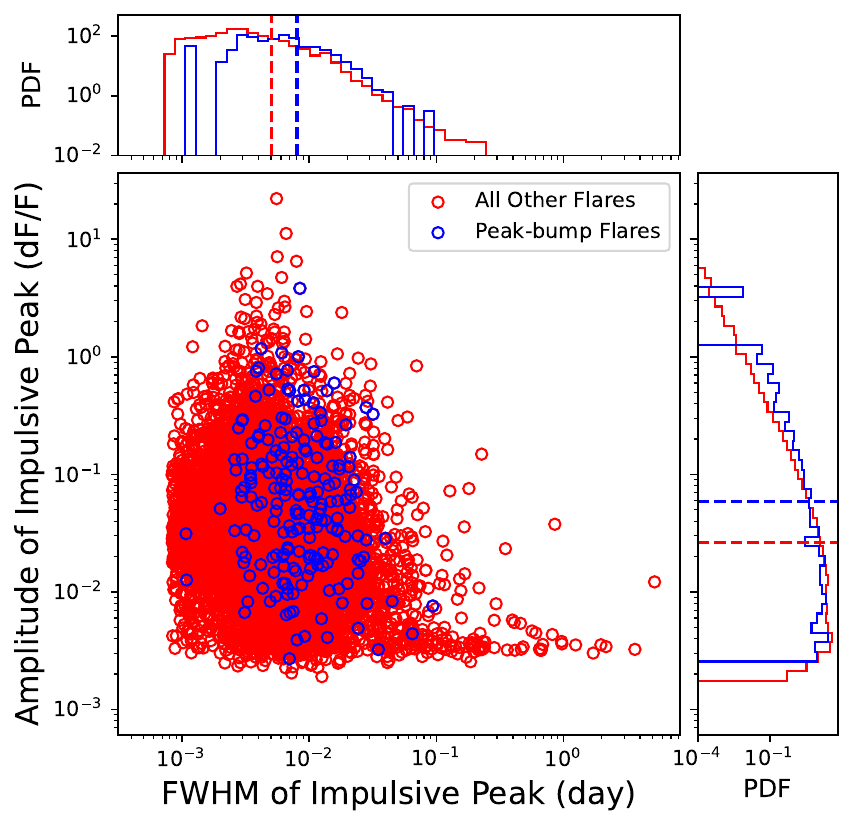}
    \caption{Shown is a comparison of the impulsive peaks' amplitudes and FWHM durations for the identified peak-bump flares (displayed as blue dots) and all other fitted flares in our samples (displayed as red dots). The marginal histograms show the probability distribution functions (PDFs) for the amplitude and FWHM values of peak-bump and all other fitted flares in our sample. The blue and red dashed lines in the PDFs represent the median values of the corresponding features for the labeled flares. }
    \label{Figure 7.}
\end{figure}

We further analyze how peak-bump flares differ from all other flares in our sample in terms of the amplitudes and FWHM values of their impulsive peaks. Figure \ref{Figure 7.} reveals that peak-bump flares (shown with the blue dots) present longer FWHM values and larger amplitudes than the other flares in our sample (shown with the red dots). This may be the result of the large flare energies associated with peak-bump flares. Since there are more large-energy superflares ($>10^{33}$ erg), relative to sample size, in our peak-bump sample (143 flares) than in our sample of all other flares (4,293 flares), having larger impulsive amplitudes and longer FWHM durations for peak-bump flares agrees with the correlations shown in Figures \ref{Figure 5.}(a) and \ref{Figure 6.}(a), respectively.

\newpage

\section{Discussions and Summary}\label{s.method5}


In this study, inspired by \cite{Howard_&_MacGregor}, we examine the occurrence of peak-bump flares on F- through M-type stars among a narrowed-down dataset comprising 10,142 flares. Utilizing seven fitting models, we aim to isolate peak-bump flares by categorizing all flares in our sample, paying closer attention to the former. Specifically, we modeled peak-bump flares by fitting a classical flare template from \cite{Davenport_et_al._2014} for the impulsive phase, combined with an empirical function (either a Gaussian, Log-normal, or Chi-squared distribution) to fit the bump. Using stringent statistical criteria and visual inspection, we identified 197 ``true" peak-bump flares. These comprise $27\substack{+4 \\ -10}$ flares from G-type stars, $51\substack{+25 \\ -15}$ flares from K-type stars, and $109\substack{+11 \\ -15}$ flares from M-type stars. This represents a conservative estimate of the occurrence rate of $1.9\%$ in our sample (10,142 flares), in contrast to the $7\%$ occurrence rate reported by \cite{Howard_&_MacGregor} based on a smaller sample of 440 flares.

Quantitative comparison of the peak and bump components presents an overall positive correlation across all stellar types. Specifically, there is a Spearman correlation coefficient of 0.69 between flare peak and bump amplitudes, and 0.33 between their FWHM durations.
While a coefficient of 0.33 indicates a weak correlation, a relationship between the two variables is vaguely seen in log-log space (Figure \ref{Figure 6.}(d)). The correlation between the amplitudes suggests that there possibly exists a related physical mechanism that drives the peak-bump flare process. The fitting of the aforementioned amplitude and FWHM relations gives a power-law index of 0.69 $\pm$ 0.09 and 1.0 $\pm$ 0.15, respectively. Additionally, we find that flares with larger energies have greater peak and bump amplitudes, as well as longer FWHM durations, refer to Figures \ref{Figure 5.} and \ref{Figure 6.}. Moreover, our analysis suggests an inverse correlation between peak-bump amplitudes and stellar effective temperatures, refer to Figure \ref{Figure 5.}(a) and (b). However, it's unclear whether this relationship is driven by underlying physical mechanisms or stems from observational bias, given that cooler stars tend to exhibit higher flare contrasts. On the other hand, we see no correlation between peak-bump FWHM durations and stellar effective temperatures. Lastly, we also observe that the amplitude and FWHM of the impulsive peak in peak-bump flares are marginally greater than those observed in all other flare types, as seen in Figure \ref{Figure 7.} from the median values and their probability distribution functions.

A  question immediately arises: what mechanism accounts for the observed correlations? Before delving into this inquiry, it's essential to reexamine flare physics and consider how insights from solar observations might contribute to our understanding.
 Solar flares that exhibit white-light emission (visible wavelength) are relatively rare but significant events. They result from the sudden release of magnetic energy stored in the solar corona, causing the rapid acceleration of waves and charged particles, including electrons, protons, and heavy ions \citep{Hudson2006SoPh..234...79H}. As these high-energy particles propagate to the lower atmosphere, they collide with atoms and ions, finally heating the photosphere and emitting white light. It is important to note that the direct collision of high-energy particles is just one hypothesis for heating the photosphere during a solar flare, and alternative mechanisms like Alfv\'{e}n wave heating and back warming by X-rays have also been proposed \citep{Emslie&Sturrock_1982, Russell_2013, Hawley&Fisher_1992}. The continuum emission in the visible spectrum can be attributed to either hydrogen free-bound emission in an enhanced region in the lower atmosphere or H-minus (H-) emission in an enhanced region around the temperature minimum \citep{Kunkel_1970,Morchenko_et_al._2015,Neidig_et_al._1993}. 

Studies of solar and stellar flares have largely concentrated on emissions from the photospheric kernel, revealing that white light flares emitted from the photosphere generally display an impulsive profile, with stronger amplitudes corresponding to more energetic flares \citep{Kowalski2017ApJ...837..125K, Kowalski2022FrASS...934458K, Hao2017}. Emissions from photospheric kernels may be spatially and temporally distinct. Additionally, multiple heating events during the decay phase could contribute to the bump phase \citep{Kosovichev&Zharkova}. However, we do not identify a quantitative relationship between the observables of these mechanisms that allows for direct comparison with the observations.

\begin{figure}
    \centering
    \includegraphics[width=0.75\linewidth]{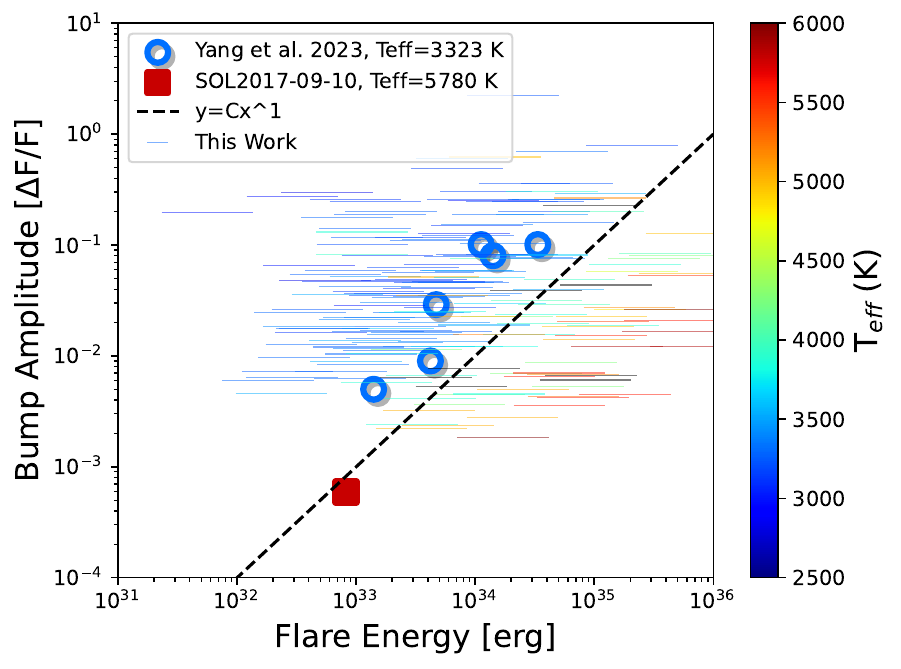}
    \caption{The horizontal lines represent data from the final peak-bump flare sample that passed the statistical tests mentioned in Section \ref{s.method3} and visual inspection with energy errors. Circles denote simulation results from \cite{Yang2023}, and the square represents data from the \textbf{SOL2017-09-10T15:35} event. The colors indicate the effective temperatures of the corresponding flare host stars. As a reference, the dashed line shows a power-law relationship with index 1.}
    \label{Figure 8.}
\end{figure}

Inspired by an extreme solar flare, \texttt{SOL2017-09-10T15:35}, which displayed an enhancement in the continuum from post flare loops \citep{Zhao2021} (note that the coronal emission not only showed a delay but also persisted for an extended duration with a gradual morphology) and a theoretical analysis of the white light continuum emission from the coronal plasma  \citep{Heinzel2017,Heinzel2018,Jejcic2018}, \cite{Yang2023} proposed another possible mechanism based on flare-induced coronal rain. This mechanism proposes that the heat from a recent flaring episode induces strong chromospheric evaporation, which strongly enhances the coronal plasma density. Then, after prolonged heating, coronal plasma condensation is triggered \citep{Parker1953, Field1965, Claes2020}, and it is the emission from the cool condensed plasma that could contribute to white light observations during the bump phase from TESS. This mechanism provides a comparable magnitude for the bump structure. Figure \ref{Figure 8.} compares the observational data collected in this study and the energy calculations from \cite{Crowley_et_al._2022} with the simulation results from \cite{Yang2023}, including the sample from \texttt{SOL2017-09-10T15:35}. The corresponding energy uncertainties are estimated by a change in the flare temperature from 8,000 to 20,000 K. These simulation results support the positive correlation found between flare energy and bump amplitude and they match the observational data points at the corresponding stellar effective temperature, implying that the coronal rain phenomenon may be a possible mechanism behind the bump in peak-bump flares, or, at the very least, the results don't exclude the coronal rain phenomenon as a possibility. Moreover, from the simulations, they found that the larger the flare energy, the stronger and longer the late phase, supporting the amplitude and FWHM correlations from this work (see Figures \ref{Figure 5.}(b) and \ref{Figure 6.}(b)). 
This model is aligned with the observations reinforced by comparing peak-bump flares with other flares, revealing that the former exhibit a longer and larger FWHM and amplitude during the initial impulsive phase, as shown in Figure \ref{Figure 7.}. It is worth noting that the flare-induced coronal rain hypothesis is just one of many to describe the late phase. Lastly, similar results were observed in the peak-bump flare sample obtained through visual inspection, without applying the strict statistical criteria mentioned in Section \ref{s.method3}, as presented in Appendix \ref{app:3}.

Upon examining the emission mechanism of the bump in simulations by \cite{Yang2023}, it is found that the rising phase is predominantly characterized by a combination of hydrogen Free-free and Free-bound emissions, with the post-peak phase of the bump being mainly dominated by Free-bound emission. 
Additionally, further surveys of stellar flares across various stellar types could refine the statistics of peak-bump flares and their feature correlations, such as energy, amplitude, and FWHM duration. More precise energy estimations of flares would further elucidate the associated dynamics, particularly through multi-wavelength observations.

\textit{Acknowledgments:} We thank James Crowley for sharing the flare catalog, and the anonymous referee and the statistics editor for improving this work.
We would like to thank the National Science Foundation (NSF) Research Experience for
Undergraduates program for funding our work (grant $\# 2050710$), the Institute for Astronomy at the University of Hawai'i, and the Department of Astronomy at the University of Florida. Xudong Sun acknowledges support from NSF award \#1848250. We also acknowledge University of Florida Research Computing for providing computational resources and support that have contributed to the research results reported in this publication. Funding for the TESS mission is provided by NASA's Science Mission directorate.

\software{lightkurve \citep{2018ascl.soft12013L}, Numpy \citep{harris2020array}, Matplotlib \citep{Hunter2007}, SciPy \citep{2020SciPy-NMeth}, Pandas \citep{McKinney_2010}}, Seaborn \citep{Waskom_2021}, and Scikit-learn \citep{2011JMLR...12.2825P}.





\newpage

\begin{appendix}
\section{QPP and Flat-top Flare Templates and Classification}
\label{app:1}
Below we outline the procedure and results of modelling all 12,597 flares in our sample with the flare templates not previously addressed: the ``QPP Model" and the ``Flat-top Model."
    
\begin{figure}
        \centering
        \includegraphics[width=0.75\linewidth]{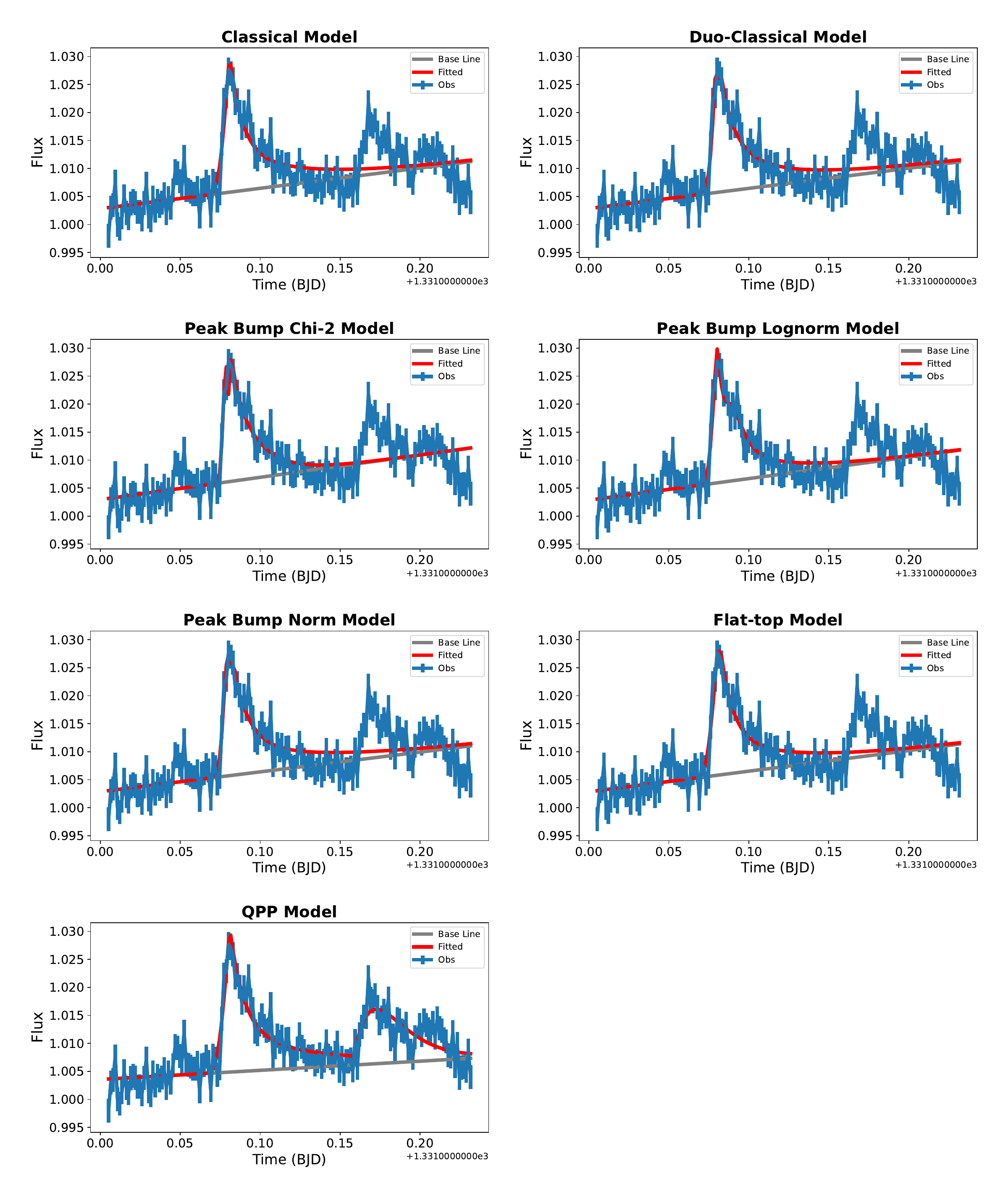}
        \caption{Example of a QPP flare identified via our algorithm. According to both the AIC and BIC criteria, this flare is best fit by the QPP Model, solidifying its QPP flare identity.
}
    \label{Figure 9.}
\end{figure}

We model QPPs by combining the classical template \citep[from][]{Davenport_et_al._2014} with a sine function multiplied by an exponential decay function to search for these small periodic fluctuations. Below, $D$ represents the coefficient corresponding to the flare's decay phase (see Equation 4 in \cite{Davenport_et_al._2014}), $a$ is the flare's start time, $d$ is the flare's decay time, $T$ is the flare's period, and $\Phi$ is the phase, all given by \cite{Crowley_et_al._2022}:
\begin{equation}
\mathrm{QPP} = De^{(a - t)/d}\sin \left( (t - a)/T + \Phi \right)
\end{equation}
QPPs were identified as those flares best fit by the QPP Model, as agreed upon by both the AIC and BIC criteria (procedure shown in Table \ref{Table 4.}). As a result, there are 95 QPPs in our sample. Figure \ref{Figure 9.} shows an example of a QPP flare fit by all seven models, emphasizing that the best fit is provided by the QPP Model, as agreed upon by both criteria.

\begin{figure}[t]
        \centering
        \includegraphics[width=0.75\linewidth]{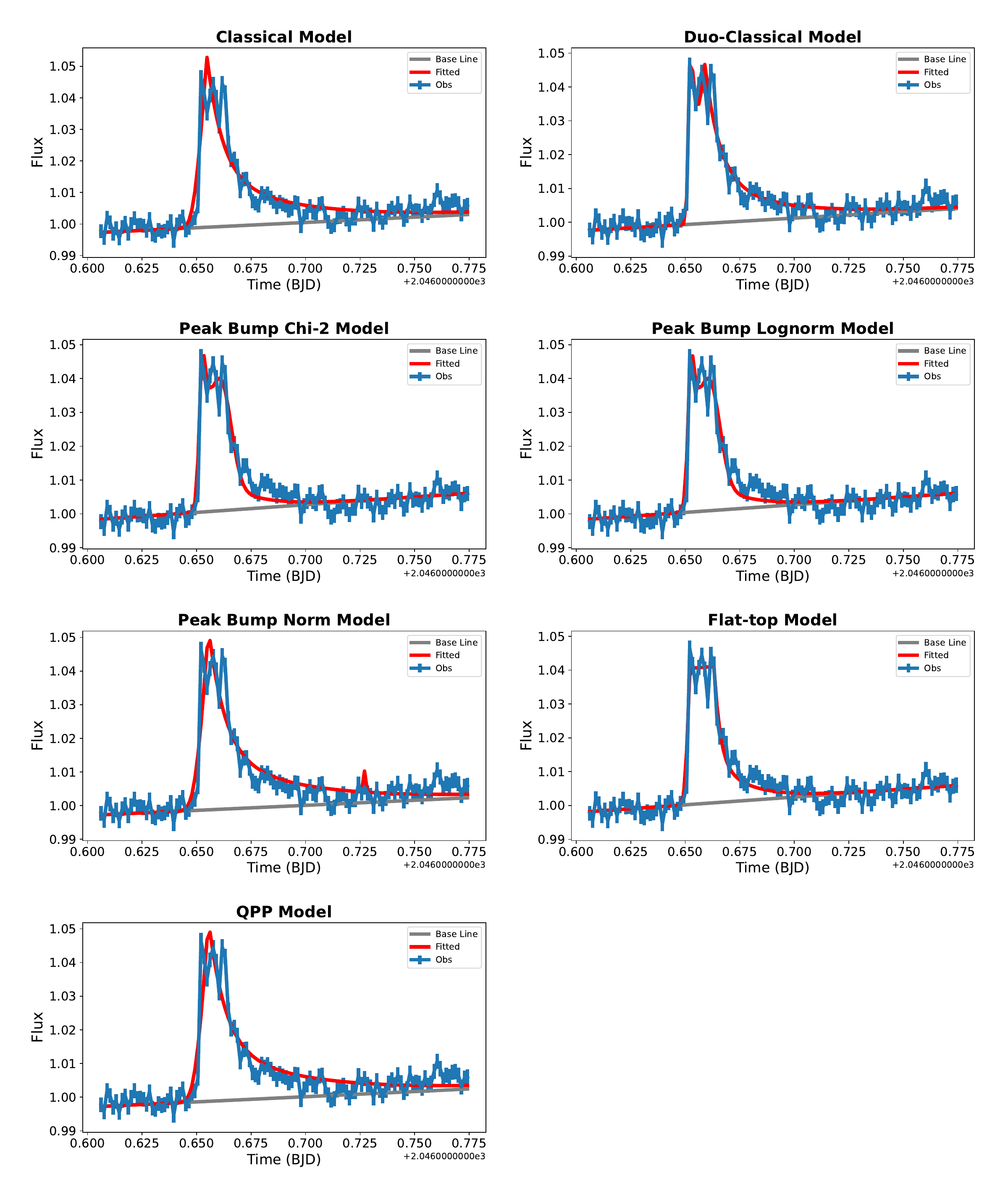}
        \caption{Example of a flat-top flare identified via our algorithm. According to both the AIC and BIC criteria, this flare is best fit by the Flat-top Model, solidifying its flat-top flare identity
}
    \label{Figure 10.}
\end{figure}

To search for flat-top flares, we modify the classical template by introducing a period of constant flux at the flare's peak that is greater than 0.8 times the FWHM. The flat-top flares are those best fit by the Flat-top Model, where both the AIC and BIC criteria have to agree on this model; in total, there are 1,277 flat-top flares in our sample (procedure shown in Table \ref{Table 4.}). Figure \ref{Figure 10.} shows an example of a flat-top flare fit by all seven models, emphasizing that the best fit is provided by the Flat-top Model, as agreed upon by both criteria.

\begin{table}[b]
\begin{center}
\caption{\label{Table 4.}Flat-top and QPP Classification Procedure}
\setlength\extrarowheight{-3pt}
\movetableright=-0.5in
\begin{tabular}{|l|l|l|}
    \hline
    \textbf{Classification} & \textbf{AIC Result} & \textbf{BIC Result}  \\ \hline
    \multirow{1}{10em} {Flat-top Flare} & Flat-top Model & Flat-top Model \\ \hline
    \multirow{1}{10em} {QPP Flare} & QPP Model & QPP Model\\ 
    \hline
    
\end{tabular}
\end{center}
\textbf{Notes.} Flares are designated as flat-tops or QPPs if both the AIC and BIC criteria conclude that the best fit model is the Flat-top Model or the QPP Model, respectively.
\end{table}

\clearpage

\section{Classical and Peak-bump Flare Classification}
\label{app:2}

\begin{figure}
    \centering
    \includegraphics[width=0.75\linewidth]{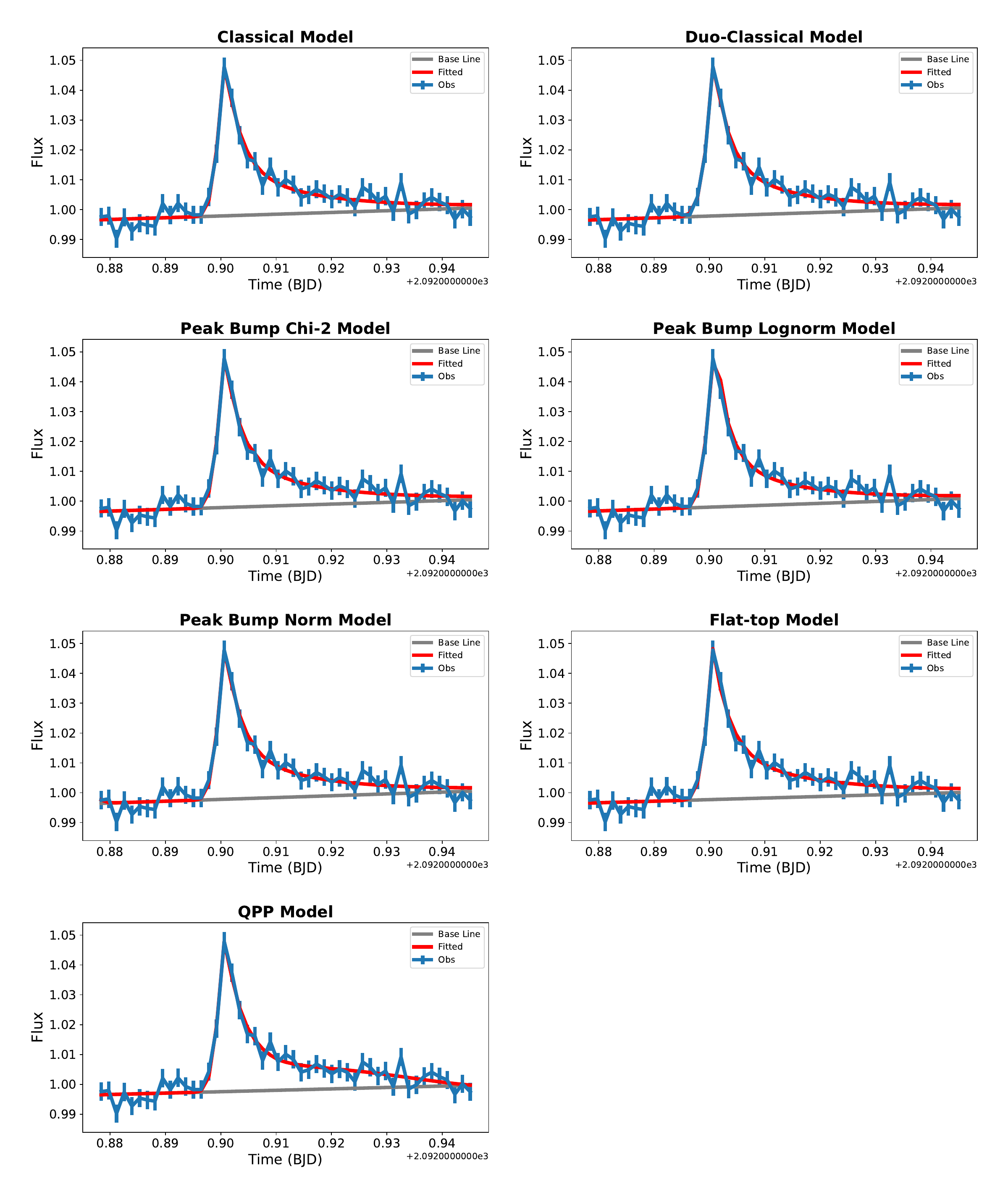}
    \caption{Example of a classical flare identified via our algorithm. According to both the AIC and BIC criteria, this flare is best fit by the Classical Model, solidifying its classical flare identity.  }
    \label{Figure 11.}
\end{figure}

We fit the flares in our sample with the flare templates mentioned in Section \ref{s.method3} and Appendix \ref{app:1} and we classified the classical and peak-bump flares based on the criteria detailed below. 
The flares that are best fit by the Classical Model are considered classical flares; both the AIC and BIC criteria must agree that the Classical Model provides the best fit for the raw data. Flares can also be identified as classical if one criterion determines the Classical Model is the best fit, while the other criterion determines that either the QPP Model provides the best fit, noting that the fitted bumps have small amplitudes---less than or equal to 10 times the median of the flux error (this could simply be noise or decay phase complexity)---or the Flat-top Model does (Appendix \ref{app:1}; the flare might have a long flux peak duration, but not long enough for a flat-top classification). An example of a classical flare identified by our algorithm is shown in Figure \ref{Figure 11.}.

Peak-bump flares are classified such that both the AIC and BIC criteria agree on one of the late-phase models, or one criterion agrees on the QPP Model described in Appendix \ref{app:1} (taking note that there is only one bump in the light curve) and the other criterion agrees on one of the late-phase models. Additional classification involves the condition that the bump of these flares has an amplitude $>$ 5 times the median of the flux error. To make certain that the classical peaks are far enough away from the bumps for such flares, we ensured that for each late-phase flare, the separation between the classical peak time and bump peak time is greater than the FWHM value for that bump. We place these restrictions on the bump's amplitude and distance from the peak, to distinguish between peak-bumps and classical flares with decay phase complexity (see \citealt{Howard_&_MacGregor}). This automatic algorithm detects 1,132 of these flares in our sample, around $11\%$ of the flares in our reduced total sample. A summary of the aforementioned flare classification procedure is shown in Table \ref{Table 5.} and an example of a peak-bump flare identified by our algorithm is shown in Figure \ref{Figure 2.}.

\begin{table}[t]
\begin{center}
\caption{\label{Table 5.}Classical and Peak-bump Classification Procedure}
\vspace*{-5mm}
\setlength\extrarowheight{-3pt}
\begin{tabular}{|l|l|l|}
    \hline
    \textbf{Classification} & \textbf{AIC Result} & \textbf{BIC Result}  \\ \hline

    \multirow{1}{10em}{Removed Flare} & ``Duo-Classical" Model & ``Duo-Classical" Model \\ \hline
   
    \multirow{4}{10em}{Classical Flare} & Classical Model & Classical Model \\
    & Classical Model & QPP Model (Small Amplitude) \\
    & QPP Model (Small Amplitude) & Classical Model \\
    & Flat-top Model & Classical Model \\ 
    & Classical Model & Flat-top Model\\ \hline
    
    \multirow{3}{10em} {Peak-bump Flare} & Chi-squared, Lognorm, or Norm Model & Chi-squared, Lognorm, or Norm Model\\
    & Chi-squared, Lognorm, or Norm Model & QPP Model (one bump) \\
    & QPP Model (one bump) & Chi-squared, Lognorm, or Norm Model \\ \hline
    
\end{tabular}
\end{center}
\textbf{Notes.}  Here, a summary of our model selection procedure is shown. There are three ways a flare can be identified as classical: 1.) both the AIC and BIC criteria agree that the Classical Model provides the best fit for that flare, 2.) one of the criteria concludes that the QPP Model provides the best fit for the flare, noting that the fitted amplitude is small (see the main text for details), while the other criterion concludes that the Classical Model provides the best fit, or 3.) one criterion identifies the Classical Model as providing the best fit while the other concludes that the Flat-top Model does. As mentioned before, all flares best fit by the ``Duo-Classical" Model (as identified by both the AIC and BIC criteria agreeing that this model provides the best fit) are removed from our sample. Lastly, peak-bump flares are identified when either 1.) both criteria agree that the Chi-squared, Log-normal, or Gaussian Model provides the best fit for the flare, or 2.) one criterion concludes that one of the aforementioned peak-bump models provides the best fit, while the other criterion concludes that the QPP Model provides the best fit, noting that there is only one bump being fitted. 
\end{table}

\clearpage

\section{Larger Visually Inspected Sample of Peak-bump Flares}
\label{app:3}

Prior to applying the statistically rigorous criteria mentioned in Section \ref{s.method3}, we had originally solely visually inspected the sample of 1,132 algorithm-detected peak-bump flares, determining a larger sample of 426 late-phase flares. The remaining flares that passed visual inspection but failed the statistical test and were thus not included in our final sample of 197, still posed merit as verified by visual inspection. Since this larger peak-bump sample provides more data and may elucidate further information about late-phase flares, below we replicate Figures \ref{Figure 5.} through \ref{Figure 8.} with this larger data sample and briefly discuss the results. 

\begin{figure}
    \centering
    \includegraphics[width=1.0\linewidth]{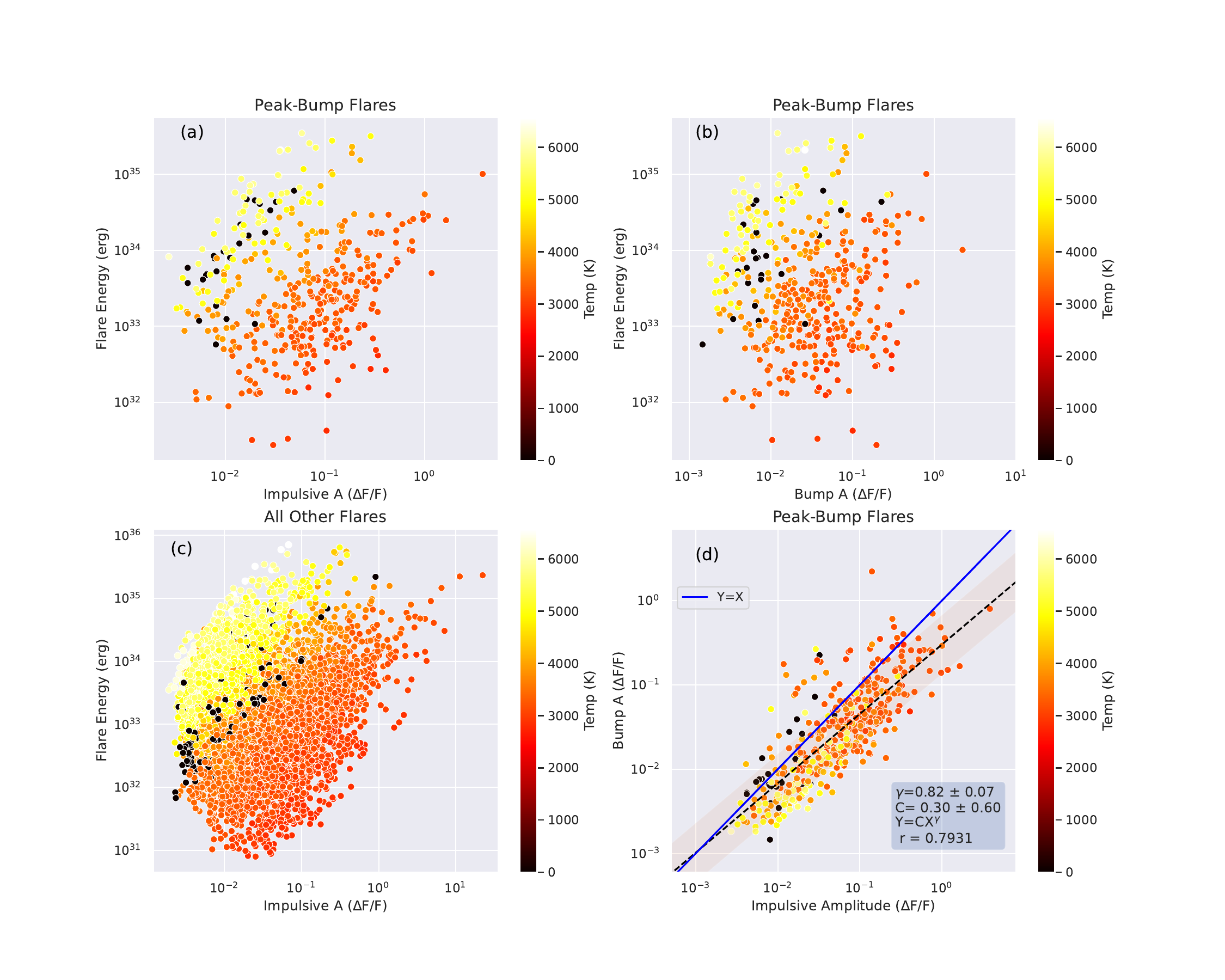}
    \caption{A comparison between the amplitudes of each fitted flare component for 426 peak-bump flares.
Plot (a) shows a positive relation between flare energy and the amplitude of the impulsive component. Plot
(b) shows a positive correlation between flare energy and the bump amplitude. Panel (c) shows a positive
relation between flare energy and the impulsive amplitude for all other flares in our sample. Lastly, panel
(d) shows the positive relation between the amplitudes of the impulsive and bump components. Overall, the
color of each dot indicates the effective temperature of the flare’s source star from the TIC v8 catalog, while
the black dots represent the stars with undetermined temperatures. The Spearman correlation coefficient of
plot (d) is 0.79. The fitted power-law index of the impulsive-bump amplitude relation in plot (d) is 0.82
± 0.07.
}
    \label{Figure 12.}
\end{figure}

\begin{figure}
    \centering
    \includegraphics[width=1\linewidth]{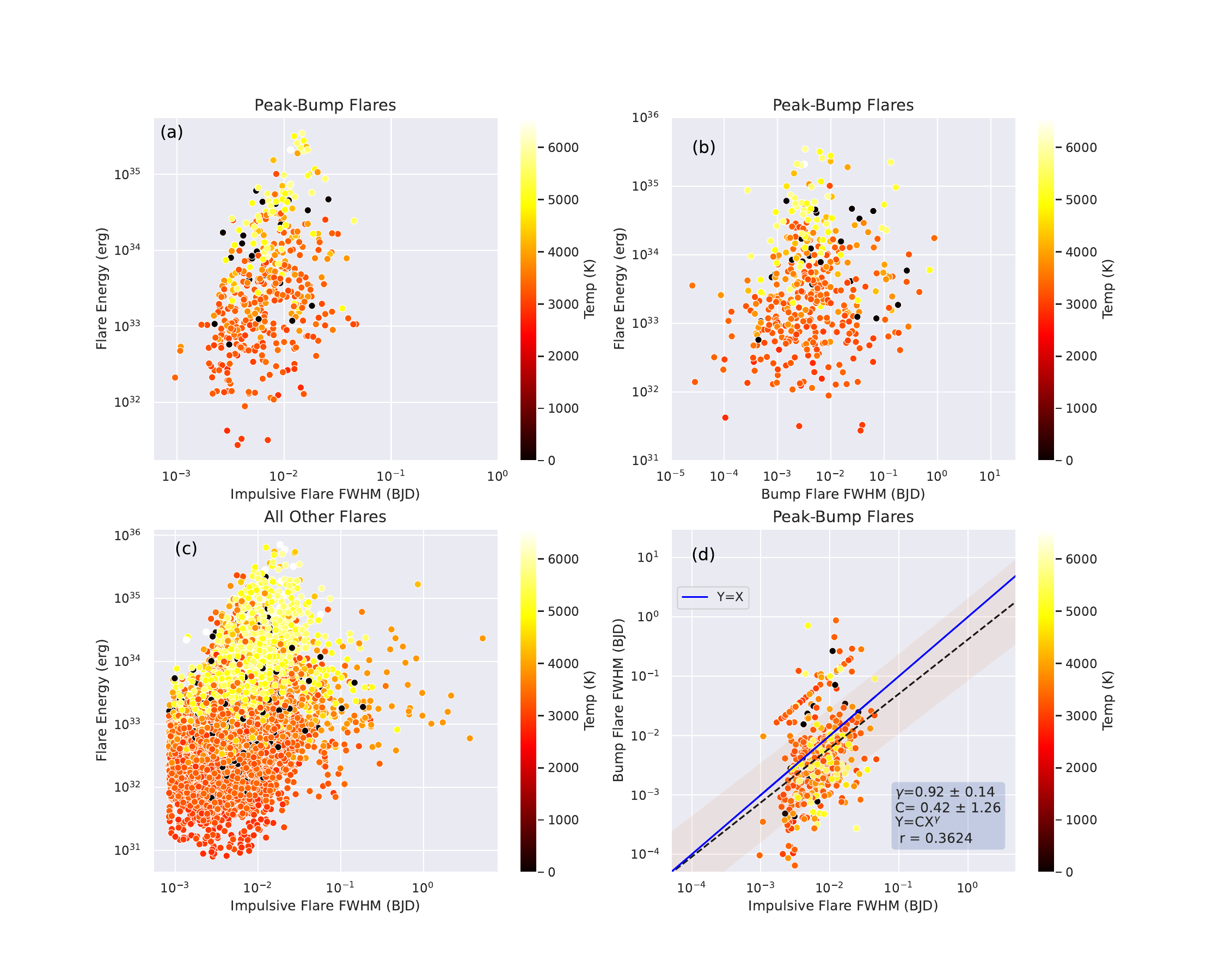}
    \caption{Same as Figure \ref{Figure 12.} but for the FWHM duration of each flare component for peak-bump flares. The Spearman correlation coefficient of panel (d) is 0.36, and the fitted power-law index of the relation in (d) is 0.92 ± 0.14. We note that the grouping of points above the fitted trendline represents the data of the late-phase flares with longer FWHM durations for their bump components than those for the flares outside the grouping. }
     \label{Figure 13.}
\end{figure}

\begin{figure}
    \centering
    \includegraphics[width=0.55\linewidth]{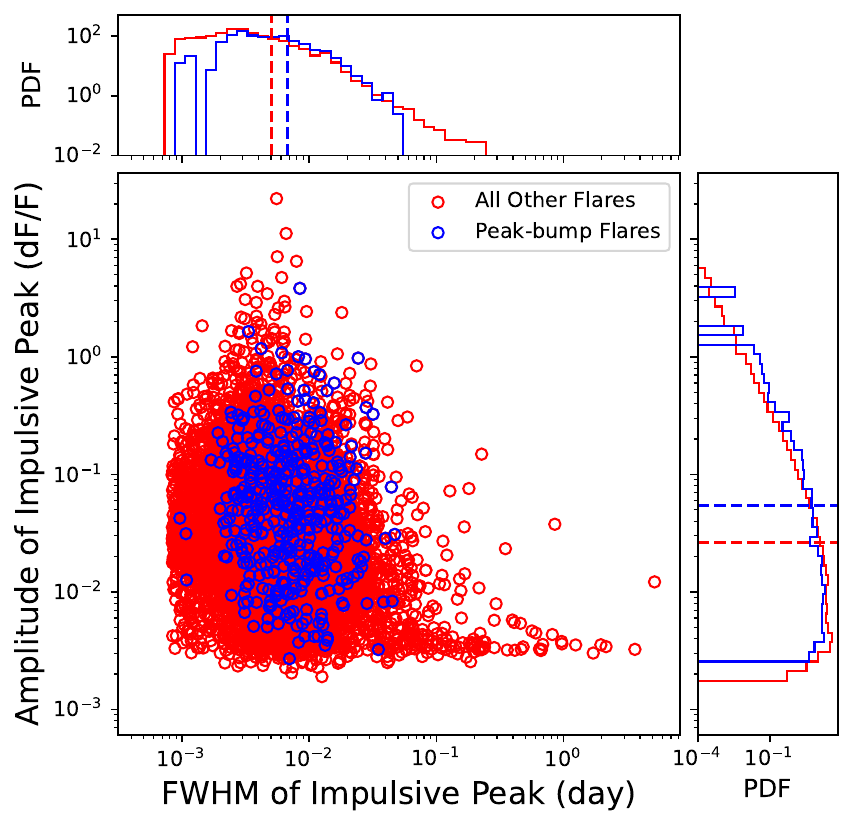}
    \caption{Shown is a comparison of the impulsive peaks' amplitudes and the FWHM durations for the identified 426 peak-bump flares (displayed as blue dots) and all other fitted flares in our larger sample (displayed as red dots). The marginal histograms show the probability distribution functions for the amplitude and FWHM values of late-phase and all other fitted flares in our sample. The blue and red dashed lines in the PDFs represent the median values of the corresponding features for the labeled flares. }
    \label{Figure 14.}

    \centering
    \includegraphics[width=0.55\linewidth]{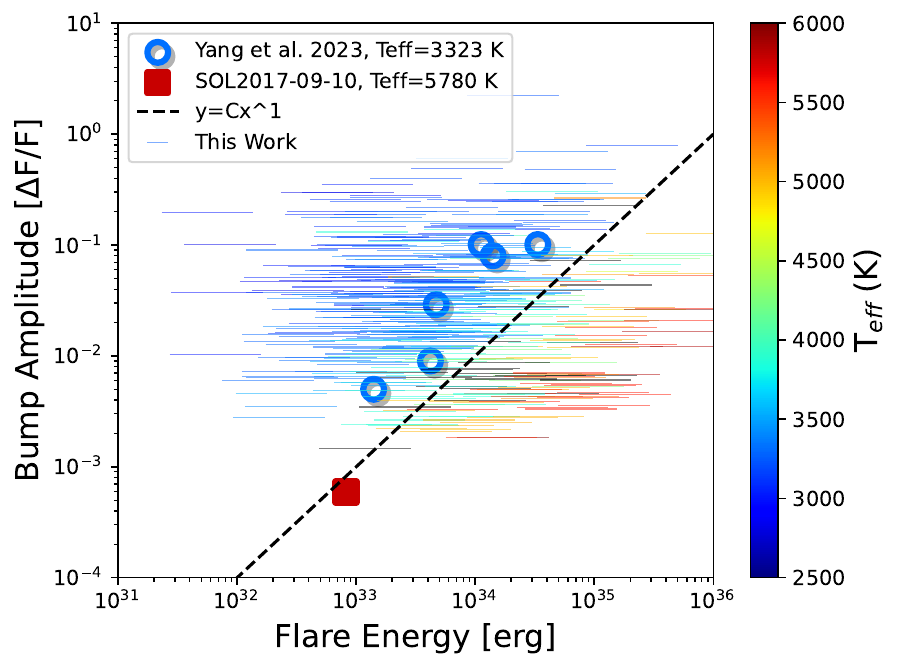}
    \caption{The horizontal lines represent data from the 426 peak-bump flare sample (that only passed visual inspection) with energy errors. Circles denote simulation results from \cite{Yang2023}, and the square represents data from the SOL2017-09-10T15:35 event. The colors indicate the effective temperatures of the corresponding flare host stars. As a reference, the dashed line shows a power-law relationship with index 1.}
    \label{Figure 15.}
\end{figure}

In Figure \ref{Figure 12.} panel (d), the Spearman correlation coefficient has increased to 0.79 and the new power law index is 0.82
± 0.07. This larger data sample provides a higher correlation coefficient than our final 197 flare sample and is thus worth mentioning. The Spearman Correlation coefficient in Figure \ref{Figure 13.}(d) is also slightly greater with a value of 0.36. The power law index for this relation is 0.92 ± 0.14. We also observe that the amplitude and FWHM of the impulsive peak in peak-bump flares are greater than those observed in
all other flare types, as seen in Figure \ref{Figure 14.} from the median values and their probability distribution functions. This agrees with Figure \ref{Figure 7.}.

Lastly, Figure \ref{Figure 15.} compares the observational data from the 426 peak-bump flare sample and the energy calculations from \cite{Crowley_et_al._2022} with the simulation results from \cite{Yang2023}. The simulation results support the positive correlation found between flare energy and bump amplitude and they match the observational data points at the corresponding stellar effective temperature, implying that the coronal rain phenomenon may be a possible mechanism behind the bump in peak-bump flares, or, at the very least, the results don’t exclude the coronal rain phenomenon as a possibility.

Overall, the mention of this larger peak-bump flare sample is important when considering the relations between certain late-phase flare properties, such as flare energy, FWHM duration, and amplitude. Although this sample did not meet the statistically rigorous criteria we set (see Section \ref{s.method3} for details), upon visual inspection we found this sample to be representative of peak-bump flares.

\end{appendix}

\clearpage

\FloatBarrier
\bibliography{sample631}{}
\bibliographystyle{aasjournal}

\end{document}